\documentclass[aps,prb,showpacs,twocolumn,superscriptaddress]{revtex4}
\usepackage{amsmath}
\usepackage{amssymb}
\usepackage{graphicx}
\begin{document}
\title{Magnetism, Conductivity and Orbital Order in (LaMnO$_3$)$_{2n}$/(SrMnO$_3$)$_{n}$ Superlattices}
\author{Shuai Dong}
\affiliation{Department of Physics and Astronomy, University of Tennessee, Knoxville, Tennessee 37996, USA}
\affiliation{Materials Science and Technology Division, Oak Ridge National Laboratory, Oak Ridge, Tennessee 32831, USA}
\affiliation{Nanjing National Laboratory of Microstructures, Nanjing University, Nanjing 210093, China}
\author{Rong Yu}
\affiliation{Department of Physics and Astronomy, University of Tennessee, Knoxville, Tennessee 37996, USA}
\affiliation{Materials Science and Technology Division, Oak Ridge National Laboratory, Oak Ridge, Tennessee 32831, USA}
\author{Seiji Yunoki}
\affiliation{Computational Condensed Matter Physics Laboratory, RIKEN, Wako, Saitama 351-0198, Japan}
\affiliation{CREST, Japan Science and Technology Agency (JST), Kawaguchi, Saitama 332-0012, Japan}
\author{Gonzalo Alvarez}
\affiliation{Computer Science and Mathematics Division and Center for Nanophase Materials Science, Oak Ridge National Laboratory, Oak Ridge, Tennessee 37831, USA}
\author{J.-M. Liu}
\affiliation{Nanjing National Laboratory of Microstructures, Nanjing University, Nanjing 210093, China}
\author{Elbio Dagotto}
\affiliation{Department of Physics and Astronomy, University of Tennessee, Knoxville, Tennessee 37996, USA}
\affiliation{Materials Science and Technology Division, Oak Ridge National Laboratory, Oak Ridge, Tennessee 32831, USA}
\date{\today}

\begin{abstract}
The modulation of charge density and spin order in (LaMnO$_3$)$_{2n}$/(SrMnO$_3$)$_n$ ($n$=$1$-$4$) superlattices is studied via Monte Carlo simulations of the double-exchange model. G-type antiferromagnetic barriers in the SrMnO$_{3}$ regions with low charge density are found to separate ferromagnetic LaMnO$_{3}$ layers with high charge density. A metal-insulator transition perpendicular to superlattices with increasing $n$ is observed, which provides insight into how disorder-induced localization may give rise to the metal-insulator transition occurring at $n$=$3$ in experiments.
\end{abstract}
\maketitle

\textit{Introduction.} Transition-metal oxide heterostructures provide a new avenue to utilize the complex properties of strongly correlated electronic materials to produce multifunctional devices. Several exotic phenomena emerge in these heterostructures due to the reconstruction at the interfaces, such as the existence of a conducting state between two insulators in LaAlO$_3$/SrTiO$_3$ and LaTiO$_3$/SrTiO$_3$.~\cite{Huijben:Nm} As one of the most representative families of strongly correlated oxide materials, the manganites can also be prepared into heterostructures with other oxides, such as cuprates, and they exhibit interesting behavior, such as orbital reconstruction.\cite{Chakhalian:Sci} 

Even without involving other oxides, manganites heterostructures can be prepared utilizing manganites with different doping, e.g. LaMnO$_3$ (LMO) and SrMnO$_3$ (SMO).\cite{Koida:Prb,Bhattacharya:Apl,Smadici:Prl,May:Prb,Bhattacharya:Prl,Adamo:Apl} These two manganites are parent compounds for the wide-band manganite La$_{1-x}$Sr$_x$MnO$_3$ (LSMO). At low temperature ($T$), bulk LaMnO$_3$ is an A-type antiferromagnetic (A-AFM) insulator, while SrMnO$_3$ is a G-type antiferromagnetic (G-AFM) insulator.\cite{Wollan:Pr} The alloy-mixed LSMO is a ferromagnetic (FM) metal at low $T$ and $0.17$$<$$x$$<$$0.5$. However, the LMO-SMO superlattices can behave differently from bulk LSMO even with the same average charge density: {\it (i)} the ordered A-site cations in the superlattices remove the A-site disorder, which is important in alloy  manganites; {\it (ii)} the artificially modulated A-site cations also modulate the physical properties, such as charge density, magnetism, and conductivity. In fact, recent experiments on (LMO)$_{2n}$/(SMO)$_{n}$ superlattices highlighted the existence of an exotic metal-insulator transition (MIT) at $n$$=$$3$.\cite{Smadici:Prl,May:Prb,Bhattacharya:Prl,Adamo:Apl} Moreover, LMO thin films on a SrTiO$_3$ (STO) substrate were found to be FM instead of A-AFM.\cite{Bhattacharya:Prl,Adamo:Apl}

Theoretically, in addition to \textit{ab-initio} calculations,\cite{Nanda:Prb} most previous model Hamiltonian investigations on manganite heterostructures were based on the one-orbital model, \cite{Kancharla:Prb} missing the important orbital degree of freedom. Although more realistic two-orbital models were used very recently,\cite{Brey:Prb,Lin:cm} several properties of the (LMO)$_{2n}$/(SMO)$_{n}$ superlattices are still not understood, particularly the explanation for the $n$=$3$ MIT.

\textit{Models and Techniques.}  The two-orbital double-exchange (DE) model is here used to study (LMO)$_{2n}$/(SMO)$_{n}$ superlattices via Monte Carlo (MC) simulations. This model Hamiltonian has been extensively studied before and it is successful to reproduce the several complex phases in manganites.\cite{Dagotto:Prp} Details about the Hamiltonian and MC technique can be found in previous publications.\cite{Dong:Prb08} Schematically, the Hamiltonian reads as:
\begin{equation}
H=H_{\rm DE}(t_0)+H_{\rm SE}(J_{\rm AF})+H_{\rm EP}(\lambda)+\sum_i(\epsilon_i-\mu)n_i,
\end{equation}
where $H_{\rm DE}$, $H_{\rm SE}$, and $H_{\rm EP}$ are the standard two-orbital large-Hund-coupling DE, superexchange (SE), and electron-phonon (EP) interactions, respectively.\cite{Dagotto:Prp} $n_i$ is the $e_{\rm g}$ charge density at site $i$. $\mu$ is the uniform chemical potential, and $\epsilon_i$ is the on-site effective potential generated by long-range Coulomb interactions that cannot be neglected in superlattices involving different electronic compositions. There are four main input parameters: the SE coupling $J_{\rm AF}$, the EP coupling $\lambda$, $\mu$, and $\epsilon_i$. All these parameters are in units of $t_0$, which is the DE hopping between nearest-neighbor (NN) $d_{3z^2-r^2}$ orbitals along the $z$ direction.\cite{Dagotto:Prp,Dong:Prb08} The constant-density phase diagram is determined by $J_{\rm AF}$ and $\lambda$. The expected $e_{\rm g}$ charge density is obtained by tuning $\mu$. Due to the valence difference between La$^{3+}$ and Sr$^{2+}$, the on-site Coulomb potential $\epsilon_i$ is inhomogeneous for the Mn sites. In almost all previous model investigations, the Coulomb interaction is treated using the Hartree-Fock (HF) approximation,\cite{Kancharla:Prb,Brey:Prb,Lin:cm} where the Poisson's equation must be solved self-consistently. However, this self-consistent process is rather difficult to converge for the three-dimensional two-orbitals model when both $t_{\rm 2g}$ classical spins and lattice distortions are also MC-time evolving. For this reason, here we adopt another strategy. Each $\epsilon_i$ will be determined by its eight NN A-site cation neighbors.\cite{Bouzerar:Prb} More specifically, in the LMO-SMO superlattices, $\epsilon_i$ is $0$ for those Mn between two LaO planes (LMO region), it becomes $V/2$ for those between LaO and SrO planes, and finally it is $V$ for those between two SrO planes (SMO region) (Fig.~1(a-b)). Therefore, this (positive) constant $V$ is the only parameter to regulate the Coulomb potential, and it is related with the dielectric constant in the HF approach. Our approximation is expected to capture the main physics in the LMO-SMO superlattices since the effective Coulomb potential is mainly caused by the modulation of A-site cations, and we believe that our qualitatively simple results shown below do not depend on these assumptions. The Coulomb screening by the $e_{\rm g}$ electron redistribution is also taken into account in part by regulating the value of $V$. In our simulation, four sets of $V$: $0.3$, $0.6$, $0.9$, and $1.2$ are used, covering the realistic potential drop range between LMO and SMO. The important practical fact is that this approximation enables the MC simulation on large enough lattices. The effort presented here would have been impossible if the Poisson's equation were to be solved iteratively during the MC evolution.

\begin{figure}
\vskip -0.3cm
\centerline{\includegraphics[width=260pt]{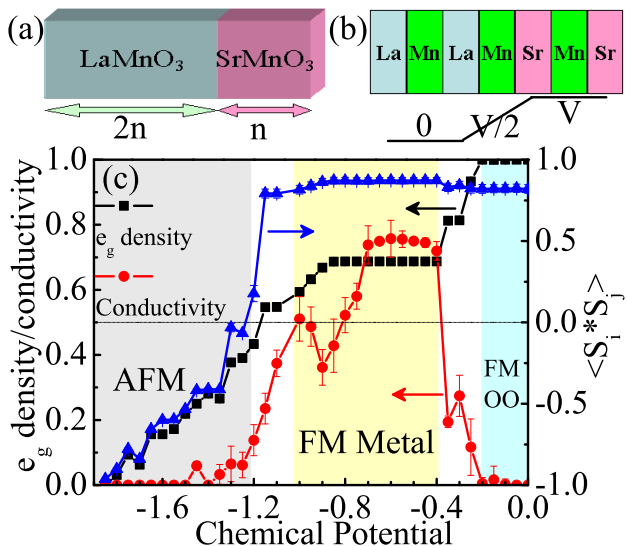}}
\vskip -0.5cm
\caption{(Color online) (a) Superlattice unit studied here. (b) On-site potential used in our simulation. (c) The $e_{\rm g}$ charge density, NN spin correlation, and conductivity (see Ref.~\onlinecite{Verges:Cpc} for its unit) of a $4$$\times$$4$$\times$$4$ lattice vs. $\mu$, at $T$=$0.01$. All $\epsilon_i$ are set to zero to simulate bulk clean-limit LSMO.}
\vskip -0.4cm
\end{figure}

For our studies, we use three-dimensional (3D, $L_x \times L_y \times L_z$) clusters with periodic boundary conditions (PBCs). Both $L_x$ and $L_y$ are set to $4$, while $L_z$ equals $3n$ ($1$$\leqslant$$n$$\leqslant$$4$). Thus, the superlattices are grown along the $z$ direction ([001]) (Fig.~1(a-b)). The MC simulation on the $4\times4\times12$ lattice is already at the cutting-edge of current computational resources. To characterize physical properties, the charge density, spin structure factors, and conductivity are calculated.\cite{Dong:Prb08,Verges:Cpc}

\textit{LaMnO$_3$: FM vs. A-AFM.} Before the simulation of superlattices, it is essential to understand why the LMO thin films on STO are experimentally found to be ferromagnetic, instead of A-AFM. For most $R$MnO$_3$ ($R$=La, Pr, Nd, Sm, and Eu), the A-AFM phase is the ground state.\cite{Zhou:Prl} However, in the previously obtained theoretical phase diagram for the two-orbitals DE model, the A-AFM regime was found to be rather narrow in parameter space, while the FM and orbital-ordered (OO) phase was clearly more robust.\cite{Hotta:Prl} Thus, to understand the FM nature of LMO thin films, we should consider lattice distortions in real manganites. In the bulk, the LMO lattice transits from a cubic perovskite to an orthorhombic one at $T\sim800$ K, below which the lattice constant along the $c$ axis shortens compared with those along the $a$ and $b$ axes.\cite{Carvajal:Prb} For instance, at $T$=$300$~K, $l_c$ is only $\sim0.964$~$l_{ab}$, where $l_c$ ($l_{ab}$) is the NN Mn-Mn distance along the $c$ axis (within the $a$-$b$ plane). Using an empirical formula by Zhou and Goodenough,\cite{Zhou:Prb} the AFM exchange intensity along $c$ becomes about $1.3$ times that on the $a$-$b$ plane. Thus, this stronger AFM coupling along $c$ will favor the A-AFM state.\cite{Hotta:Prl} However, for LMO thin films on STO, the LMO lattice is compressed in the $a$-$b$ plane but it is elongated along the $c$ axis, leading to an almost cubic crystal structure.\cite{Bhattacharya:Apl} Therefore, the theoretical phase diagram,\cite{Hotta:Prl} derived assuming lattice isotropy, should be applicable to the LMO thin films that prefer the FM/OO phase instead of the A-AFM one.

Following the theoretical phase diagram,\cite{Hotta:Prl} here we choose a particular set of parameters ($J_{\rm AF}$=$0.09$, $\lambda$=$1.2$) for the simulation below. To justify this choice, first we perform a MC calculation on a $4\times4\times4$ lattice (PBCs) with all $\epsilon_i=0$ to examine whether the above parameters are suitable for LSMO. The $e_{\rm g}$ charge density $n_i$, NN spin correlation ($<S_i\cdot S_j>$), and conductivity are calculated at low $T$ ($T$=$0.01$, $\sim60$ K if $t_0\sim0.5$ eV), as shown in Fig.~1(c). For $n_i\sim1$, the phase is found to be FM but insulating, in agreement with the results found for experimental LMO thin films. The energy gap at the Fermi level is $0.3$ ($\sim 150$ meV), also in agreement with the experimental excitation energy ($\approx125$ meV).\cite{Bhattacharya:Prl} The insulating character of the state is caused by orbital ordering driven by Jahn-Teller distortions, although the orbital ordering in strained cubic LMO is a little different from the well-known $d^{3x^2-r^2}$/$d^{3y^2-r^2}$ pattern in bulk.\cite{Hotta:Prb} In the density range $0.5<n_i<0.8$, the system is FM and conducting, also in agreement with the LSMO properties at the corresponding dopings. For $n_i<0.5$, the NN spin-correlations turn out to be negative, suggesting an AFM phase. The conductivity becomes poor with decreasing $n_i$ until it reaches an insulating state. These $n_i<0.5$ behavior is also compatible with LSMO at the corresponding doping. Then, as a conclusion, the  set $J_{\rm AF}$=$0.09$ and  $\lambda$=$1.2$ should be a proper parameter set to describe (cubic) LSMO. In the following, we will use this set to study the (LMO)$_{2n}$/(SMO)$_n$ superlattices.

\begin{figure}
\vskip -0.3cm
\centerline{\includegraphics[width=260pt]{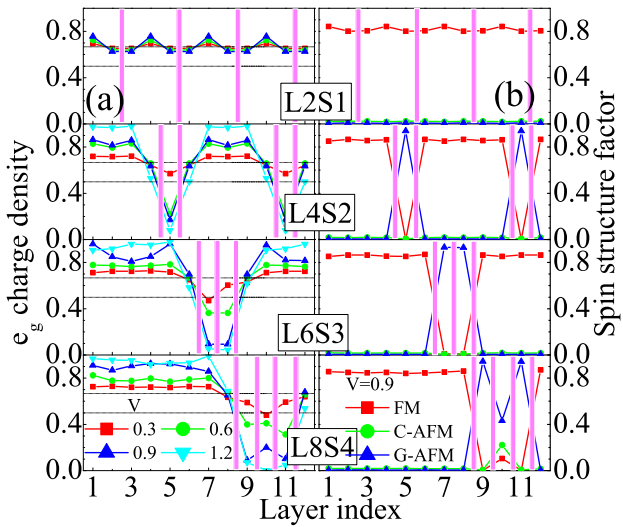}}
\vskip -0.4cm
\caption{(Color online) (a) Charge modulation in the superlattices with different $V$'s (the case of L2S1 with $V$=$1.2$ can not been obtained due
to phase separation). The two horizontal lines denote $0.5$ and $2/3$. (b) In-plane spin structure factor for $V$=$0.9$. In (a) and (b), pink bars denote SrO layers in the (LMO)$_{2n}$/(SMO)$_n$ superlattice, while LaO layers are not highlighted. The lattices are simply repeated along the $z$ direction if their periods are shorter than $12$.}
\vskip -0.4cm
\end{figure}

\textit{(LMO)$_{2n}$/(SMO)$_n$ superlattices.}
We will focus first on the low-$T$ ($=0.01$) MC results which can shed light on the ground state properties. The local $e_{\rm g}$ charge densities of each layer vs. layer index are shown in Fig.~2(a). For (LMO)$_2$/(SMO)$_1$ (L2S1), the charge distribution is fairly uniform despite the use of substantial values for $V$ ($0.3-0.9$), namely the local charge density fluctuates weakly around the average $2/3$. This is easy to understand since there is no SMO region in the L2S1 superlattice and, thus, the potential fall here is only $V/2$. For other superlattices, and usually ($V>0.3$), the densities in the SMO regions are lower than $0.5$. The $V$=$1.2$ case already restricts most of the $e_{\rm g}$ electrons to be in the LMO regions, while $V=0.3$ is low enough that it  spreads $e_{\rm g}$ electrons to the SMO regions. Therefore, it is reasonable to conclude that the potential amplitude range used in our model is the proper one to cover the potential drop in real manganites. In the following, we will focus on the case $V$=$0.9$, for which the results are similar to the experimental data.

Figure~2(b) shows the $x$-$y$ plane spin structure factors vs. layer index, for $V$=$0.9$ and $T$=$0.01$. There are only two main components: 
FM and G-AFM. The latter exists only in the SMO regions, while FM dominates in the LMO regions, and at the LMO/SMO interfaces. This spin arrangement 
agrees with experiments.\cite{May:Prb,Bhattacharya:Prl} Our calculated spin order supports the idea that the local phases in superlattices are mainly determined by the local densities $n_{\rm local}$:\cite{Lin:cm} if $n_{\rm local}>0.5$, the spin order is FM, and it is G-AFM when $n_{\rm local}$$\approx$$0$. Therefore, other spin orders, e.g. A-AFM or C-AFM, may emerge once the $n_{\rm local}$  is slightly lower than $0.5$. In fact, we observed the coexistence of several complex spin orders at, e.g., the SMO region ($n_{\rm local}$$\approx$$0.15$) for L8S4 with $V$=$0.9$ (Fig.~2(b)).

\begin{figure}
\vskip -0.3cm
\centerline{\includegraphics[width=260pt]{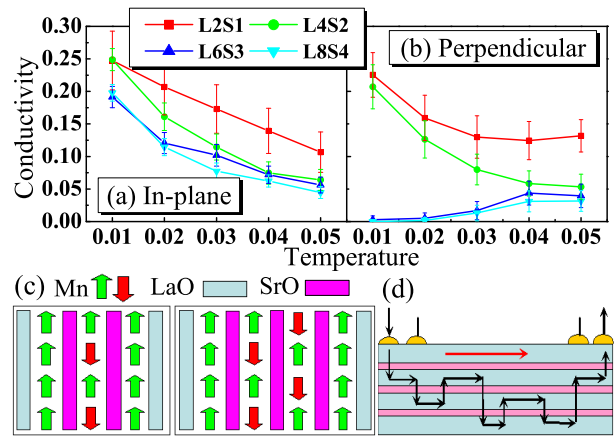}}
\vskip -0.4cm
\caption{(Color online) (a) In plane conductivity for the superlattices studied here. (b) Perpendicular conductivity. 
(c) Sketch of the spin order at interfaces for L4S2 (left) and L6S3 (right). (d) Sketch of experimental setup for resistance measurements. 
Pink bars are SMO regions. Typical conducting paths via the DE process (black curves) connect NN interfaces, but they will be broken when $n\geqslant3$.}
\vskip -0.4cm
\end{figure}

\textit{Understanding the critical n for the MIT.} 
One of the most important experimental discoveries in the (LMO)$_{2n}$-(SMO)$_n$ superlattices is the MIT with increasing $n$. To try to understand this phenomenon, here two conductivities (only from the DE process) are calculated vs. $T$: the in-plane one (along $x$ or $y$ direction) and the perpendicular one (along $z$ direction), as shown in Figs.~3(a) and (b). All in-plane conductivities are robust and increase with decreasing $T$, suggesting metallicity. Our result agrees with previous studies showing that the charge transfer at interfaces between Mott/band insulators can generate conducting interfaces.\cite{Huijben:Nm,Kancharla:Prb} In contrast, the perpendicular conductivities show metallic behavior when $n\leqslant2$, but insulating behavior when $n\geqslant3$. To understand this MIT, the spin arrangement at the FM/G-AFM interface should be considered. The intra-layer NN spin correlations (not shown here) confirm that spins at the FM/G-AFM interfaces are almost collinear at low $T$, namely the NN spins are parallel or antiparallel, as shown in Fig.~3(c). Therefore, when there is only one G-AFM layer in each superlattice unit ($n$=$2$), the ``green'' spin-up channels of the G-AFM layer link the NN FM layers, allowing for a good conductance. However, once the SrO layers thickness is $3$, the two G-AFM layers cut off the same-spin channels, giving rise to an insulating behavior along the $z$ direction. Therefore, it is natural to expect that $n$=$3$ must be the MIT critical point in (LMO)$_{2n}$/(SMO)$_n$ superlattices.

However, it should be noted that the experimental resistance measurements were performed using the four-point method (Fig.~3(d)).\cite{May:Prb,Bhattacharya:Prl} Thus, to fully understand the experimental MIT, Anderson localization effects should also be taken into consideration. When $n\geqslant3$, as mentioned before, the G-AFM insulating barriers cut down conduction channels along the $z$ direction. Thus, the DE conducting process becomes exclusively 2D-like (i.e. only within the $x-y$ plane). In this case, insulating behavior is likely induced by interface disorder, such as roughness, and Anderson localization, as already discussed in the experimental literature.\cite{Bhattacharya:Prl} But a 3D metallicity in the $n\leqslant2$ superlattices, avoiding the Anderson localization mechanism, can be achieved by the DE process perpendicular to the interfaces, as shown schematically in Fig.~3(d). In short, although the in-plane conductivity is metallic in our small lattices without disorder, the critical $n$ for the MIT in the real superlattices still corresponds to the same value found in our simulations via studies of the perpendicular conductivity.

\begin{figure}
\vskip -0.3cm
\centerline{\includegraphics[width=260pt]{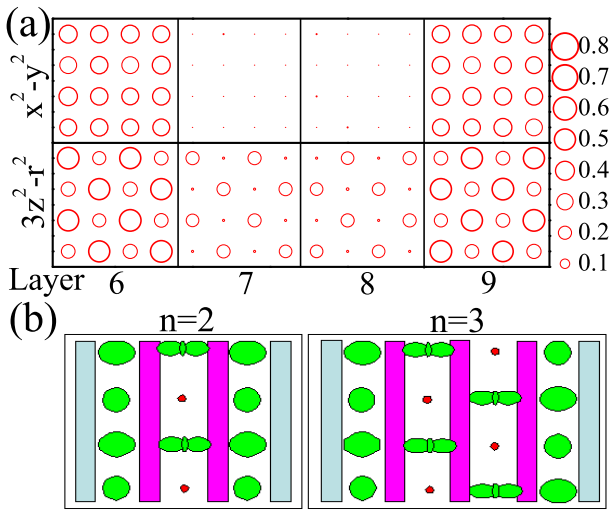}}
\vskip -0.4cm
\caption{(Color online) (a) Orbital occupation of the $6$th-$9$th layers for L6S3 when $V=0.9$ and $T=0.01$. The $6$th and $9$th are interface layers between the LaO and SrO layers, while the $7$th and $8$th are within the SMO region. Here the circle's area is proportional to the local $e_{\rm g}$ charge density. (b) Sketch of orbital ordering at the interface for $n$=$2$ and $3$.}
\vskip -0.4cm
\end{figure}

\textit{Orbital order.} 
The spin configuration in Fig.~3(c) can also induce orbital order near the interfaces. It should be pointed out that the strain/stress 
due to the lattice mismatch between substrates and LMO/SMO can induce orbital order even without charge transfer.\cite{Nanda:Prb} 
However, here we propose that the the parallel spin channels (in both the uncut and cut cases of Fig.~3(c)) can provide another driving force 
for orbital order in manganite superlattices. Our simulations show in the $n$=$2$, $3$, $4$ cases, that the $e_{\rm g}$ electrons have more tendency 
to occupy the $d_{3z^2-r^2}$ orbital than the $d_{x^2-y^2}$ one, to improve the kinetic energy in the growth direction of the superlattice.
One example is shown in Fig.~4. This orbital order optimizes the DE process between the FM regions via the G-AFM layers. But in real superlattices
the combined effect of this spin-driven tendency to orbital order and that induced by strain/stress may compete and more quantitative calculations will
be needed to decide which orbital order dominates.

\textit{Conclusions.} We have performed a MC simulation to study the two-orbitals double-exchange model for the (LMO)$_{2n}$-(SMO)$_n$ superlattices. First, we have explained why the LMO thin films on STO are FM instead of A-AFM. Then, our simulations have shown that the spin order in the SMO regions is G-AFM, while it is FM elsewhere. The spin arrangement between the FM and G-AFM layers causes the metal-insulator transition, with $n$=$3$ as the critical value.

We thank A. Bhattacharya, S. May, M. Daghofer, and S. Okamoto for helpful discussions. Work was supported by the NSF grant DMR-0706020 and the Division of Materials Science and Engineering, U.S. DOE, under contract with UT-Battelle, LLC. S.Y. was supported by CREST-JST. G.A. was supported by the CNMS, sponsored by the Scientific User Facilities Division, BES-DOE. J.M.L. was supported by the 973 Projects of China (2006CB921802) and NSF of China (50832002). S.D. was supported by the China Scholarship Council.

\end{document}